\documentstyle[12pt,epsf,axodraw]{article}
\title{ {Triple and Quartic Interactions of Higgs Bosons in the
                 General Two-Higgs-Doublet Model}}

\author{
         M.N.Dubinin \\
       {\small \it Institute of Nuclear Physics, Moscow State
University}\\  \vspace{2mm}
       {\small \it  119899 Moscow, Russia} \\ 
         A.V.Semenov \\
       {\small \it Joint Institute for Nuclear Research} \\ 
       {\small \it Dubna, Moscow Region, Russia}  }
\date{}

\begin{document}
\maketitle

\begin{abstract}
In the case of minimal supersymmetric extension of the Standard
Model (MSSM), when the pseudoscalar Higgs boson mass is less than
the supersymmetry energy scale, the effective theory at
the electroweak scale is a two-Higgs-doublet model. We diagonalize the 
mass matrix of the general two-Higgs-doublet model, expressing
Higgs boson self-couplings in terms of two mixing angles and 
four Higgs boson masses, and derive in a compact form the complete set of
Feynman rules, including quartic couplings in the Higgs sector, for
the case of $CP$-violating potential.
Some processes of double and triple Higgs boson production at a 
high-energy linear collider are calculated in the case of 
mixing angles and scalar boson masses satisfying the MSSM constraints.
\end{abstract}

\newpage

\section{Introduction}

A particularly simple extension of the Standard Model containing
two scalar doublets \cite{Georgi} has been very extensively
investigated in the framework of minimal supersymmetry. In order
to cancel gauge anomalies introduced by the fermionic superpartners
of gauge bosons and to generate masses of up- and down-
quarks in a consistent manner two doublets of Higgs fields are
necessary. 

Soft supersymmetry-breaking terms \cite{softSB} introduce large
radiative corrections to the tree-level Higgs boson masses and
couplings \cite{RC} and the effective lagrangian of the Higgs sector
at the electroweak scale does not satisfy the supersymmetry constraints
valid at the SUSY scale. In the most general case when the supersymmetry
scale and the scale of heavy Higgs boson mass (usually defined by
the mass of the pseudoscalar) are different, the effective theory
at the electroweak scale is a two-Higgs-doublet model, where the
self-interaction couplings are defined by the renormalization
group evolution of the supersymmetric potential couplings from the SUSY 
scale down to the electroweak scale \cite{RGevol,HaberHempfling}.

The investigation of direct phenomenological consequences of a 
two-doublet Higgs sector at a future high luminosity colliders, such as 
LHC and TESLA, could provide a possibility to study in detail the
structure of effective Higgs potential, mass spectrum and couplings of the
scalar particles. As usual, the variety of channels where scalars
could be produced individually or in association with vector bosons
requires a systematical calculation in order to find out what
particular channels could have a sufficient counting rates for
experimental detection at a given collider luminosity.

We propose a convenient compact form of Feynman rules for a
general two-Higgs-doublet model that can be used in the following
systematical study of the Higgs boson production channels and use
these rules for the calculation of two and three Higgs boson
production at a high energy $e^+ e^-$ collider.

\section{Diagonalisation of the mass matrix in the general
         two-Higgs-doublet model}

General form of the (nonsupersymmetric) $SU(2)\times U(1)$ invariant
potential in the case of two doublets of complex scalar fields
$\varphi_1$, $\varphi_2$ can be found in \cite{HHG}

\begin{eqnarray}
V(\varphi_1,\varphi_2)=&
         \lambda_1 (\varphi_1^+ \varphi_1 -\frac{v_1^2}{2})^2
        +\lambda_2 (\varphi_2^+ \varphi_2 -\frac{v_2^2}{2})^2 \\ \nonumber
       & +\lambda_3 [(\varphi_1^+ \varphi_1 -\frac{v_1^2}{2})
                  +(\varphi_2^+ \varphi_2 -\frac{v_2^2}{2})]^2 \\
\nonumber
       & +\lambda_4 [(\varphi_1^+ \varphi_1)(\varphi_2^+ \varphi_2)
                  -(\varphi_1^+ \varphi_2)(\varphi_2^+ \varphi_1)] \\ \nonumber
 &  +\lambda_5[Re(\varphi_1^+ \varphi_2)-\frac{v_1 v_2}{2} Re(e^{i\xi})]^2
   +\lambda_6[Im(\varphi_1^+ \varphi_2)-\frac{v_1 v_2}{2} Im(e^{i\xi})]^2
\end{eqnarray}
where $\lambda_i$ are real constants.
Components of scalar doublets $\varphi_{1,2}$ are
\begin{equation}
\varphi_1 = \{ -iw^+_1, \frac{1}{\sqrt{2}}(v_1+h_1+iz_1)\}, \quad
\varphi_1 = \{-iw^+_2, \frac{1}{\sqrt{2}}(v_2+h_2+iz_2)\}.
\end{equation}
where $w$ is a complex field and $z$, $h_{1,2}$ are real scalar fields.
Vacuum expectation values $v_1$, $v_2$ correspond to the minimum of the
potential
\begin{equation}
\varphi_1= \frac{1}{\sqrt{2}} \{ 0, v_1 \}, \quad
\varphi_2= \frac{1}{\sqrt{2}} \{ 0, v_2 e^{i \xi} \}
\end{equation}
where the phase $\xi$ can be removed by the rotation of $\varphi_1^+
\varphi_2$ not affecting the $\lambda_4$ term in (1).
Substiution of (2) to (1) gives a bilinear form of the mass term with
mixed components $w,h_{1,2},z$, which can be diagonalized by an orthogonal
transformation of the fields in order to define the tree level masses of
physical bosons. The resulting spectrum of scalars consists of two charged
$H^\pm$, three neutral $h$, $H$, $A^0$ scalar fields, and three Goldstone
bosons $G$. This procedure is described in many papers (for instance,
\cite{HHG, GHI}). The $w_{1,2}$ sector is diagonalized by the rotation of
$w_1, w_2 \rightarrow H,G$
\begin{equation}
w_1^\pm =-H^\pm s_\beta +G^\pm c_\beta, \quad
w_2^\pm = H^\pm c_\beta + G^\pm s_\beta
\end{equation} 
defined by the angle
\begin{equation}
{\tt tg} \beta = \frac{v_2}{v_1}
\end{equation}
and leading to the massless $G$ field and the field of massive charged Higgs
boson $H^\pm$, $m_{H\pm}^2=\lambda_ 4(v_1^2+v_2^2)/2$. 
The $z_{1,2}$ sector is diagonalized by the rotation 
$z_1,z_2 \rightarrow A^0, G^{'}$ defined by
the angle $\beta$ (4) and giving again one massless field $G^{'}$ and the 
field of CP-odd Higgs boson $A^0$ with the mass 
$m_{A}^2=\lambda_5(v_1^2+v_2^2)/2$.
Finally, the $h_1,h_2$ sector
is diagonalised by the rotation $h_1, h_2 \rightarrow h, H$ defined by the
angle $\alpha$ 
\begin{equation}
{\tt sin} 2\alpha =\frac{2m_{12}}{\sqrt{(m_{11}-m_{22})^2+4m_{12}^2}},
\quad
{\tt cos} 2\alpha
=\frac{m_{11}-m_{22}}{\sqrt{(m_{11}-m_{22})^2+4m_{12}^2}}
\end{equation}
where
\begin{eqnarray*}
m_{11}=&\frac{1}{4}[4v_1^2(\lambda_1+\lambda_3)+v_2^2\lambda_5] \\ \nonumber
m_{22}=&\frac{1}{4}[4v_2^2(\lambda_2+\lambda_3)+v_1^2\lambda_5] \\
\nonumber
m_{12}=&\frac{1}{4}(4\lambda_3+\lambda_5) v_1 v_2
\end{eqnarray*}    
giving two massive fields of CP-even Higgs bosons $H,h$ with the
mass values
\begin{equation}
m^2_{H,h}=m_{11}+m_{22} \pm \sqrt{(m_{11}-m_{22})^2+4m_{12}^2}
\end{equation}
In the explicit form the diagonal mass matrix of scalar fields and the
physical boson interaction vertices are obtained after the following
substitution of
$\lambda_{i}$ to the potential $V(\varphi_1, \varphi_2)$ (1):
\begin{eqnarray}
\lambda_1=&\frac{1}{2v^2} \frac{1}{c_{\beta}^2}
         [\frac{s_{\alpha}}{s_{\beta}} c_{\alpha-\beta} m_h^2
        - \frac{c_{\alpha}}{s_{\beta}} s_{\alpha-\beta} m_H^2]
        + \frac{c_{2\beta}}{4c^2_{\beta}} \lambda_5  \\ \nonumber
\lambda_2=&\frac{1}{2v^2} \frac{1}{s_{\beta}^2}
         [\frac{c_{\alpha}}{c_{\beta}} c_{\alpha-\beta} m_h^2
        + \frac{s_{\alpha}}{c_{\beta}} s_{\alpha-\beta} m_H^2]
        - \frac{c_{2\beta}}{4s^2_{\beta}} \lambda_5  \\ \nonumber
\lambda_3=&\frac{1}{2v^2}[-\frac{s_{2\alpha}}{s_{2\beta}} m_h^2
                  + \frac{s_{2\alpha}}{s_{2\beta}} m_H^2]  
                  - \frac{1}{4}\lambda_5  \\ \nonumber
\lambda_4=&\frac{2}{v^2}m_{H^\pm}^2 \\        \nonumber
\lambda_6=&\frac{2}{v^2}m_{A^0}^2                        
\end{eqnarray} 
where we used the notation $v^2=v_1^2+v_2^2$, $s_{\alpha}={\tt sin}
\alpha$,
$c_\alpha= {\tt cos} \alpha$. Diagonalization of the mass term
takes place if $\lambda_5$ is arbitrary, but the necessary
condition for the $CP$-invariance of potential (1) is 
$\lambda_5=\lambda_6$ \cite{HHG}. Unfortunately, after the substitution
of (8) to the potential (1) the intermediate expressions for the four
scalar boson interaction vertices turn out to be extremely cumberous and
it is very difficult to reduce them to some compact convenient form,
where the dependence of the coupling from the parameters could be clearly
seen. This is a technical problem of the symbolic manipulaion program
\cite{LanHEP} that we used. However, symbolic transformations of the
intermediate expressions
are simpler, if we rewrite the potential $V(\varphi_1, \varphi_2)$
in the oftenly used representation 
\begin{eqnarray}
U(\varphi_1, \varphi_2)=& -\mu_1^2 (\varphi_1^+ \varphi_1)
                         -\mu_2^2 (\varphi_2^+ \varphi_2)
       -\mu_{12}^2 (\varphi_1^+ \varphi_2+ \varphi_2^+ \varphi_1) \\  \nonumber     
     &+\bar \lambda_1 (\varphi_1^+ \varphi_1)^2
     +\bar \lambda_2 (\varphi_2^+ \varphi_2)^2
   +\bar \lambda_3 (\varphi_1^+ \varphi_1)(\varphi_2^+ \varphi_2) \\  \nonumber
   &+\bar \lambda_4 (\varphi_1^+ \varphi_2)(\varphi_2^+ \varphi_1)
    +\frac{\bar \lambda_5}{2} 
       [(\varphi_1^+ \varphi_2)(\varphi_1^+\varphi_2)
          + (\varphi_2^+ \varphi_1)(\varphi_2^+ \varphi_1)]
\end{eqnarray}
It is easy to check that in the case of zero $\varphi_1^+ \varphi_1$ phase
the potentials (1) and (9) are equivalent if the constants 
$\bar \lambda_i$, $\mu$ and $\lambda_i$ are related by the formulas
\begin{eqnarray}
\bar \lambda_1= \lambda_1+\lambda_3, \quad
\bar \lambda_2= \lambda_2+\lambda_3, \quad
\bar \lambda_3= 2\lambda_3+\lambda_4, \quad    \\ \nonumber
\bar \lambda_4= -\lambda_4+ \frac{\lambda_5}{2}+\frac{\lambda_6}{2}, \quad
\bar \lambda_5= \frac{\lambda_5}{2}-\frac{\lambda_6}{2} 
\end{eqnarray}
and
\begin{equation}
\mu_{12}^2=\lambda_5 \frac{v_1 v_2}{2}, \quad
\mu_1^2=\lambda_1 v_1^2+\lambda_3 v_1^2 +\lambda_3 v_2^2, \quad
\mu_2^2=\lambda_2 v_2^2+\lambda_3 v_1^2 +\lambda_3 v_2^2
\end{equation}                        
The expressions (11) are sometimes called 'minimization conditions', if
one starts from the potential $U(\varphi_1,\varphi_2)$ (9), where the
symbolic structure does not show clearly a possible minimum.
In the MSSM $\bar \lambda_5=$0 and it follows that $\mu^2_{12}$ is
fixed and equal to $m^2_A s_{\beta} c_{\beta}$. If this equality
is not satisfied (or, equivalently, $\lambda_5 \neq \lambda_6$ in (1)), 
$CP$-violation in the Higgs sector can be introduced.
The diagonal form of $U(\varphi_1,\varphi_2)$ and the physical scalar
boson interaction vertices are obtained by the substitution of the
following expressions for $\bar \lambda_{i}$ and $\mu_i$ in the
potential (9):
\begin{eqnarray}
\bar \lambda_1=&\frac{1}{2v^2} 
         [(\frac{s_{\alpha}}{c_{\beta}})^2 m_h^2
        + (\frac{c_{\alpha}}{c_{\beta}})^2 m_H^2  
     -  \frac{s_{\beta}}{c^3_{\beta}}\mu_{12}^2  ] \\ \nonumber
\bar \lambda_2=&\frac{1}{2v^2} 
         [(\frac{c_{\alpha}}{s_{\beta}})^2 m_h^2
        + (\frac{s_{\alpha}}{s_{\beta}})^2 m_H^2
     -  \frac{c_{\beta}}{s^3_{\beta}}\mu_{12}^2  ] \\ \nonumber
\bar \lambda_3=&\frac{1}{v^2}[2m^2_{H^\pm}   
                - \frac{\mu_{12}^2}{s_{\beta} c_{\beta}}
              -\frac{s_{2\alpha}}{s_{2\beta}} (m_H^2-m_h^2)]\\ \nonumber
\bar \lambda_4=&\frac{1}{v^2}(\frac{\mu_{12}^2}{s_{\beta} c_{\beta}} 
                   +m^2_A- 2 m^2_{H^\pm} ) \\ \nonumber
\bar \lambda_5=&\frac{1}{v^2} (\frac{\mu_{12}^2}{s_{\beta} c_{\beta}}
                    -m^2_{A} )               \\ \nonumber
\mu^2_1=&\frac{1}{2}[\frac{s_{\alpha}}{c_{\beta}}s_{\alpha-\beta}m^2_h
                   + \frac{c_{\alpha}}{c_{\beta}}c_{\alpha-\beta}m^2_H
   - 2tg\beta \, \mu_{12}^2]\\ \nonumber
\mu^2_2=&\frac{1}{2}[-\frac{c_{\alpha}}{s_{\beta}}s_{\alpha-\beta}m^2_h
                   + \frac{s_{\alpha}}{s_{\beta}}c_{\alpha-\beta}m^2_H
   - 2ctg\beta \, \mu_{12}^2]
\end{eqnarray}
Our expressions for $\bar \lambda_4$ and $\bar \lambda_5$ are
the same as given in \cite{HaberHempfling} for the case of 
zero $\lambda_6$ and $\lambda_7$.
Complete sets of Feynman rules (unitary gauge) for the triple and quartic
Higgs boson interactions in the general two-Higgs-doublet model with
a possibility of $CP$-violation in the Higgs sector (defined by $\mu_{12}$
parameter),
are shown in Tables 1-2. These sets were obtained
by means of LanHEP package \cite{LanHEP}. LanHEP package is a 
specialized symbolic manipulation system capable to generate Feynman 
rules for the $SU(2)$, $SU(3)$ gauge invariant lagrangians with arbitrary
sets of particle multiplets,
in the standard input lagrangian format of CompHEP package \cite{CompHEP}.
We do not show here a rather long set of Feynman rules in the
'tHooft-Veltman gauge, that can be also generated after the introduction
of ghost and ghost-goldstone lagrangian terms to LanHEP program.  
\footnote{The generation process takes 15 sec. of CPU time (i686).
Complete lagrangian tables in CompHEP format and LanHEP
package are available at 
{\tt http://theory.npi.msu.su/\~{}semenov/lanhep.html}}

We assume that in the Yukawa sector $<\varphi_1>$
couples only to down fermions 
\begin{equation}
V_{ud} \frac{e m_d}{2 \sqrt{2} m_W s_W c_{\beta}}
       [\bar \psi_1 (1+\gamma_5) \psi_2 \varphi_1
        + \bar \psi_2 (1-\gamma_5) \psi_1 \varphi_1^+]
\end{equation}
(here for the $u$, $d$ quarks $\bar \psi_1= \{ \bar u, V_{ud} \bar d
+V_{us}\bar s+ V_{ub} \bar b\}, \quad \psi_2=d$ and analogous structures for 
$s$,$b$ quarks and leptons, in the case of quarks $V_{ab}$ denotes the CKM
matrix elements), and $<\varphi_2>$ couples only to up fermions (so-called 
model of type II \cite{typeII}):
\begin{equation}
    \frac{e m_u}{2 \sqrt{2} m_W s_W s_{\beta}}
       [\bar \psi_1 (1+\gamma_5) i \tau_2 \psi_2 \varphi_2^+
        + \bar \psi_2 (1-\gamma_5) i \tau_2 \psi_1 \varphi_2]
\end{equation}
(here $\bar \psi_1= \{ \bar u, V_{ud} \bar d +V_{us} 
\bar s+ V_{ub} \bar b\}, \quad \psi_2=u$ and analogous structures for $c$   
and $t$ quarks).
Higgs-gauge boson interaction is defined by the straightforward
extension of the covariant derivative in the case of two scalar
doublets. It is easy to find the relation 
between the vacuum expectation values of the potential and the $W$-boson 
mass and coupling $g=e/{\tt sin} \vartheta_W$
\begin{equation}
v^2=v^2_1+v^2_2=\frac{4 m^2_W}{e^2}s^2_W
\end{equation}
following from the structure of scalar fields kinetic term 
$D_{\mu}\varphi D^{\mu}\varphi$.  

From the phenomenological point of view the general multiparametric
two-Higgs-doublet model is too flexible to be systematically used 
for data analysis.
Practically no limits on the masses of individual scalars can be set 
if their couplings to gauge bosons and fermions depend on some free
parameters, and can be very small in a rather large
regions of parameter space. Recent discussion of the possible limits 
can be found in \cite{Krawczyk}. However, the parameter space can be
strongly restricted by the constraints imposed by the supersymmetry.   
 
Let us consider the reduction of the general two-doublet model Feynman
rules shown in Tables 1,2 to the case of minimal supersymmetry model
(MSSM). The potential $V(\varphi_1, \varphi_2)$ (1) contains eight
parameters: two VEV's  $v_1$, $v_2$ and six $\lambda_i$ ($i$=1,...6).
Eight parameters of the potential $U(\varphi_1, \varphi_2)$ (9)
$\mu_1$, $\mu_2$, $\mu_{12}$ and $\bar \lambda_{i}$ ($i$=1,...5) can be
found using (10),(11). From the other side, in order to define the
Higgs sector we need eight physical parameters:
the mixing angle $\beta$ and $W$-boson mass $m_W$, mixing angle
$\alpha$, the parameter $\mu_{12}$ and four masses of
scalars $m_h$, $m_H$, $m_A$, $m^\pm$. Two VEV's can be expressed
through $m_W$, ${\tt tg}\beta$ by (5) and (15) and only one degree of
freedom remains here. In 
the case of superpotential five additional constraints are imposed,
relating all Higgs boson self couplings $\bar \lambda_{i}$, ($i$=1,...5)
to the gauge coupling constants at the energy scale $M_{SUSY}$
\cite{Inoue}:
\begin{equation}
\bar \lambda_1= \bar \lambda_2=\frac{g^2+g^2_1}{8}, \quad
\bar \lambda_3=\frac{g^2-g^2_1}{4},           \quad
\bar \lambda_4=-\frac{g^2}{2},                \quad
\bar \lambda_5=0.
\end{equation}
As we already noticed, if $\bar \lambda_5=$0, $\mu_{12}$ is
fixed and $CP$-parity is conserved. 
The remaining two independent parameters may be used to
define all Higgs boson masses and mixing angles. One can
choose, for instance, $r_1, r_2$ parametrization \cite{GHII}
($r_{1,2}=m^2_{h,H}/m^2_Z$) or the well-known $m_A$, ${\tt tg} \beta$
parametrization. In order to reduce the general two-Higgs-doublet model
vertices to the case of MSSM it is convenient to use the
$\alpha$, $\beta$ parametrization:
\begin{eqnarray}
m^2_h=m^2_Z 
{\tt cos} 2\beta \, \frac{{\tt sin}(\alpha+\beta)}{{\tt
sin}(\alpha-\beta)},
\quad
m^2_H=m^2_Z 
{\tt cos} 2\beta \, \frac{{\tt cos}(\alpha+\beta)}{{\tt
cos}(\alpha-\beta)},
\\
\nonumber
m^2_A=m^2_Z \frac{{\tt sin} 2(\alpha+\beta)}{{\tt sin} 2(\alpha-\beta)},
\quad
\mu^2_{12}= m^2_A {\tt sin} \beta {\tt cos} \beta.
\end{eqnarray}
Substitution of these expressions to the vertex factors in Tables 1,2
after trivial trigonometric transformations reduces
them to simpler MSSM factors \cite{HHG}. Complete list of Feynman
rules at the MSSM scale is shown in Table 3.

Renormalization group (RG) evolution of the coupling constants $\lambda_i$
from the energy scale $M_{SUSY}$ to the electroweak scale $M_{EW}$ 
violates the constraints (16) \cite{RC} and the effective low
energy potential at the scale $M_{EW}$ is the potential of a general 
two-Higgs doublet model with RG evolved couplings $\bar \lambda_i$.
At a given values of $m_A$, ${\tt tg}\beta$ (or $\alpha$, $\beta$), masses
of Higgs bosons and the mixing angle $\alpha$ (or $m_A$) at the energy
scale $M_{SUSY}$ can be obtained using (17). Detailed analysis of the
following RG evolution and the calculation of leading-logarithmic
radiative corrections to the mixing angles, masses and couplings of Higgs 
bosons can be found, for instance, in \cite{HaberHempfling}. We briefly
point out that the additional input parameters to be defined in order
to fix the scheme are the scale of SUSY breaking $M_{SUSY}$, the
mass parameter in higgsino-gaugino sector $\mu$, and the squark mixing
parameters $A$.

\section{Multiple production of neutral Higgs bosons}

The processes of multiple neutral Higgs boson production in the MSSM
were considered in \cite{Zerwas1, Osland} in the framework of effective
potential approach to the calculation of radiatively corrected scalar
masses and couplings \cite{Zerwas2} of the SUSY Higgs sector. The
reactions 
\begin{equation}
e^+ e^- \to hhZ, \quad e^+ e^- \to hhA, \quad e^+ e^- \to \nu_e \bar
\nu_e hh
\end{equation}
were considered and it was shown that the cross sections of double
and triple Higgs boson production are not small and the experimental
measurements of triple Higgs boson couplings are realistic.

We used the results of \cite{HaberHempfling, HMSUSY} to calculate the
radiatively corrected masses of Higgs bosons and mixing angle $\alpha$
in the renormalization group approach to the Higgs
potential couplings evolution from the SUSY scale $M_{SUSY}$ down 
to the electroweak scale (see also \cite{Zerwas3}). We set the 
$M_{SUSY}=$1 TeV and have not
included the effects of squark mixing by setting the parameters
$A$ and $\mu$ equal to zero. In the case of not too large ${\tt tg} \beta$ 
(we used ${\tt tg} \beta= 3$) and the pseudoscalar 
mass $m_A$ of order 150--250 GeV, masses of heavy CP-even Higgs boson 
$H$ and charged Higgs boson $H^{\pm}$ are also at the scale 150-250 GeV. 
The lightest Higgs boson mass is approaching 100 GeV when the pseudoscalar
mass surpasses 200 GeV. (Changes of the SUSY scale and mixing parameters
can in principle shift $m_h$ by about 50 GeV, see the details in
\cite{HaberHempfling, Osland}). These radiatively
corrected parameters were used in our set of Feynman rules. The following 
calculation of the complete tree level amplitude for the multiple Higgs
boson production processes (18) was performed by means of CompHEP
package \cite{CompHEP}, when the exact symbolic result for the matrix
element squared is converted to FORTRAN code and integrated by
multichannel Monte-Carlo method. The s-channel resonant peaks of the 
amplitude (see Fig.1) are regularized by phase space mappings \cite{IKP}
to ensure an efficient application of VEGAS integrator \cite{VEGAS}.

While in the Standard Model the cross section of $hhZ$ production
\cite{SM} is 
of order $2 \cdot 10^{-1}$ fb at the Higgs boson mass 100 GeV and
slowly
decreasing when the mass of Higgs boson increases, the picture
in the two-doublet MSSM sector is strongly changed by the 
availability of resonant production mechanisms, when the decays
of on-shell $H\to hh$ and $A^0\to Zh$ become possible.
We show the dependence of total cross sections in the channels (18)
from the masses of CP-even states $m_h$ and $m_H$
in Fig.2,3. In order to understand qualitatively the cross section
behaviour we show also the $h$, $H$ branching ratio dependence 
(in the two-body decay channels with the contribution greater than 1\%)
from their masses in Fig.4,5. Rapid decrease of the total rate at $m_h=$
60 GeV ($m_H=$ 120 GeV) and rapid increase at $m_h=$ 95 GeV ($m_H=$ 190
GeV) are directly connected with the resonant threshold of the
heavy scalar decay $H \to hh$ (see diagrams in Fig.1). The channel $e^+
e^- \to hhZ$ receives some enhancement at $m_h=$ 95 GeV ($m_H=$ 210 GeV)
when the resonance threshold $A^0 \to Zh$ is opened. 

Our results are qualitatively consistent with the results of
\cite{Zerwas1, Osland, Zerwas3}, where somewhat different regions of
the two-Higgs-doublet model parameter space were explored.
Radiatively corrected masses are rather sensitive to the input
parameter values. At smaller value of ${\tt tg} \beta$ a mass interval 
between the closing and opening $hh$ thresholds decreases to a few
GeV (\cite{Zerwas1}, ${\tt tg} \beta=$ 1.5).

The reactions (18) do not include quartic Higgs boson interaction
vertices. We calculated the cross-section of the simplest process   
\begin{eqnarray}
e^+ e^- \to hhhZ \nonumber
\end{eqnarray}
(see Fig.2,3), where quartic vertices $hhhh$ and $hhhH$ participate
(21 diagrams in the unitary gauge).
In a very limited region of parameter space the reaction
has an observable cross-section if the luminosity is high, and
the experimental reconstruction of multijet events is very efficient.

\section{Conclusions}

Large increase of the estimate of the possibly achievable integrated
luminosity in the next linear colliders (especially L$=$ 
500 fb$^{-1}$/year for the TESLA project) makes quite realistic
the experimental study of Higgs boson self-interaction. Such an
investigation is especially interesting if the Higgs sector of the
model includes more than one $SU(2)$ multiplet. Untrivial spectrum of
scalars leads to the resonant multiple Higgs boson production mechanisms,
when the final states with 4 or 6 $b$-jets from their decays will appear
with the cross sections of one-two orders of magnitude greater than 
in the SM case of only one scalar boson in the Higgs sector.

For a systematical study of various production channels we derive in
a compact form a complete set of Feynman rules for the general case of
two-Higgs doublet model. We demonstrate that in the case of minimal
supersymmetry, when additional constraints are imposed on the general
parameter space, the interaction vertices are reduced to the well-known
vertices of the MSSM at the scale $M_{SUSY}$. Useful connection
of LanHEP output in the standard lagrangian format of CompHEP input makes
possible the following efficient calculation of various reactions. 

\begin{center}
{\large \bf Acknowledgements}
\end{center}
M.D. is grateful to S.Y.Choi and P.M.Zerwas for useful discussions. The
authors would like to express their gratitude to H.S.Song and the
Center for Theoretical Physics, Seoul National University, where this 
work was completed, for hospitality. The authors acknowledge thankfully
discussions with the members of Minami-Tateya group (KEK, Tsukuba) and 
N.Skachkov. The research was partially supported by 
RFBR grant 96-02-19773a, INTAS grant INTAS-YSF 98-50 and
St.-Petersburg University KCFE grant.

\newpage

\unitlength=1cm

\begin{figure}[t]
\begin{center}      
\begin{picture}(8,8)
\put(-6.5,-15){\epsfxsize=24cm
         \epsfysize=24cm \leavevmode \epsfbox{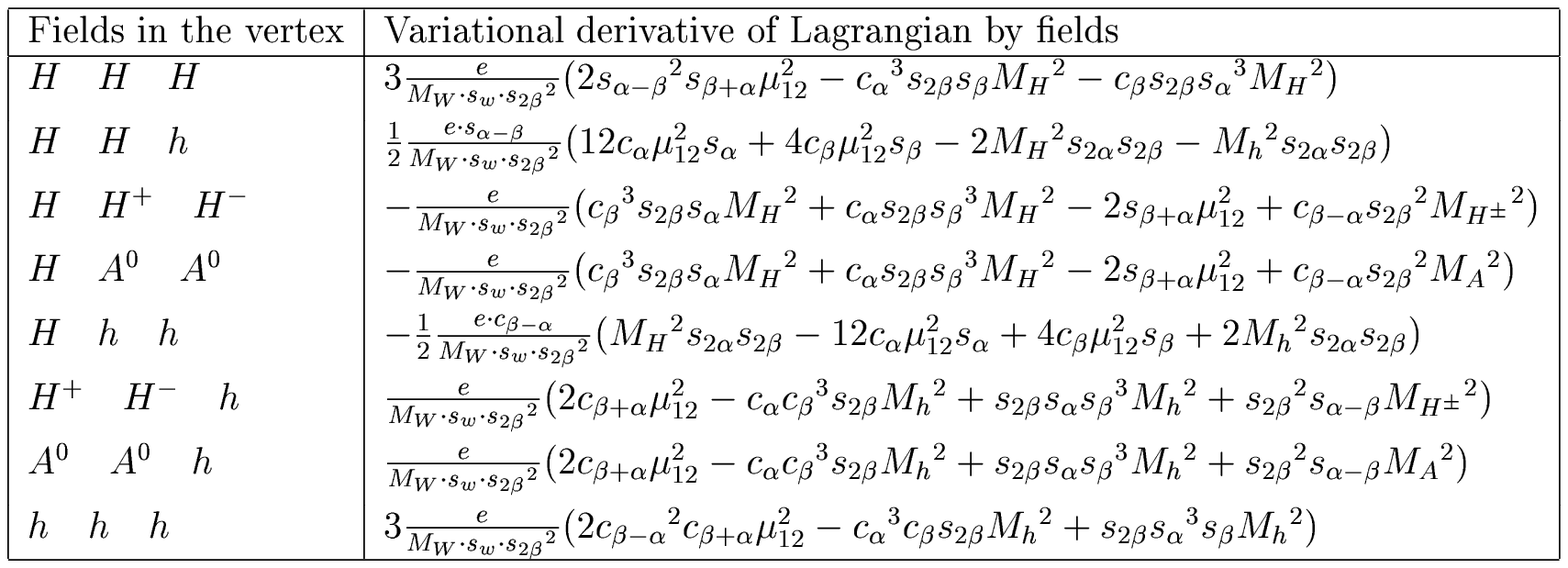}}
\end{picture}   
\end{center}  
\end{figure}

\begin{center}
 Table 1. Triple Higgs boson interaction vertices in the
general two Higgs doublet model
\end{center}

\newpage

\begin{figure}[t]
\begin{center}   
\begin{picture}(15,13)
\put(-4,-4){\epsfxsize=21cm
         \epsfysize=23cm \leavevmode \epsfbox{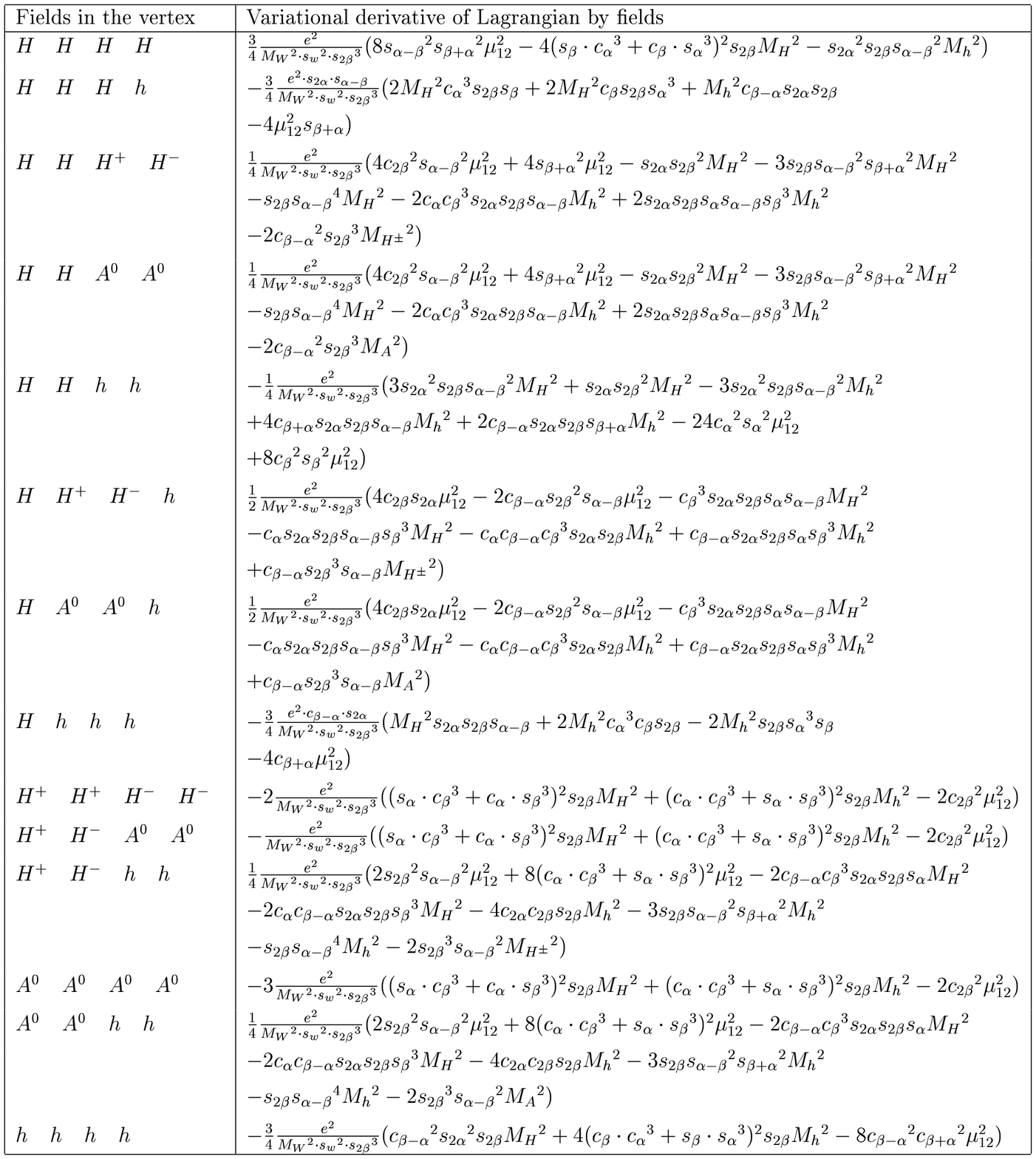}}
\end{picture}  
\end{center}
\end{figure}
.
\vskip 2cm
\begin{center}
 Table 2. Quartic Higgs boson interaction vertices in the general
two Higgs doublet model
\end{center}

\newpage

\begin{figure}[t]
\begin{center}
\begin{picture}(15,13)
\put(-2.5,-5){\epsfxsize=22cm
         \epsfysize=23cm \leavevmode \epsfbox{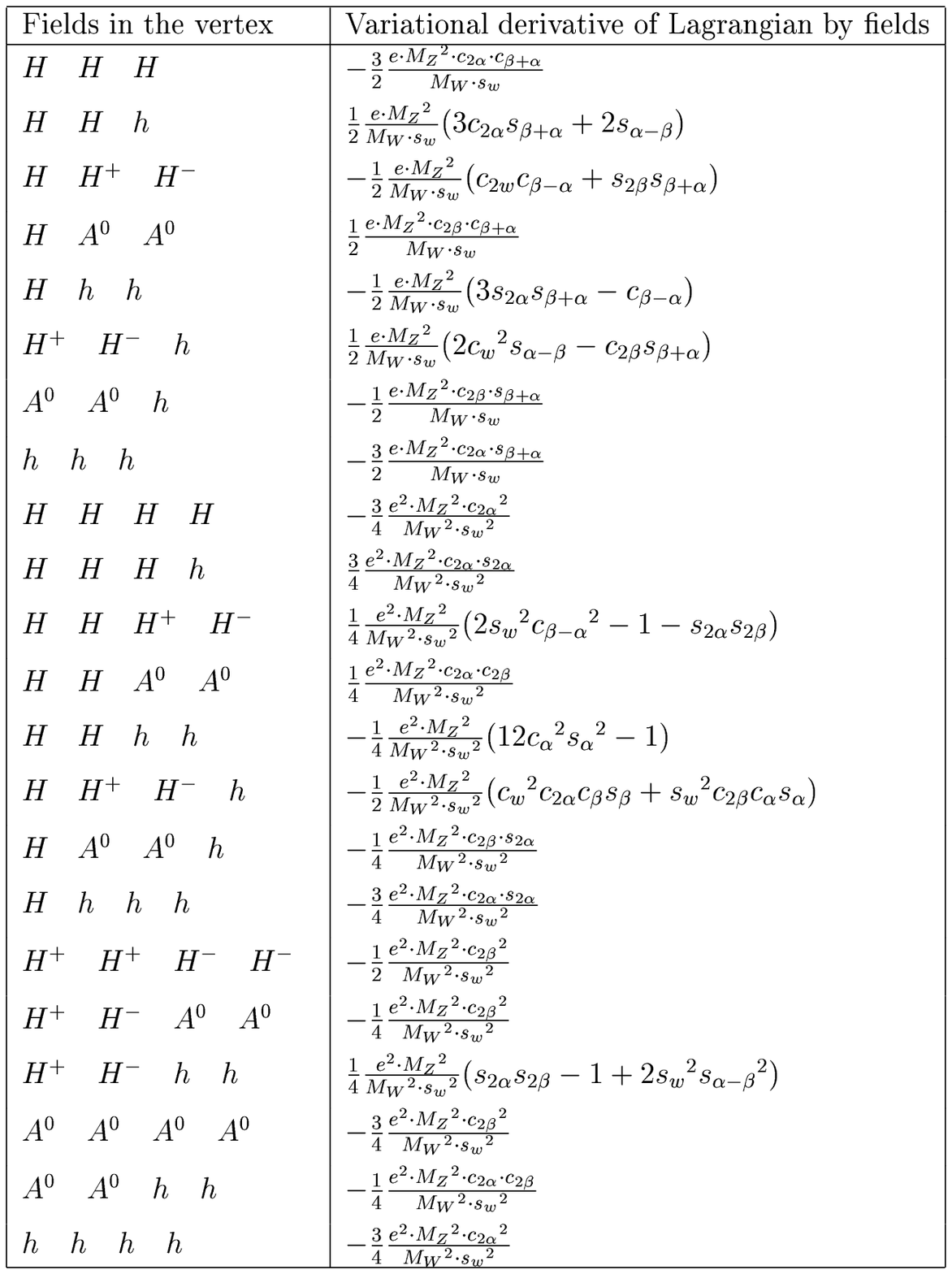}}
\end{picture}
\end{center}
\end{figure} 
\begin{center}
 Table 3. Triple and quartic Higgs boson interaction vertices at the
scale $M_{SUSY}$ 
\end{center}

\newpage

\begin{center}
{\bf Figure captions}
\end{center}

Fig. 1 Feynman diagrams for the process $e^+ e^- \to hhZ$
\vskip 5mm
Fig. 2 Total cross sections for the reactions $e^+ e^- \to hhZ$,
$e^+ e^- \to hhA$, $e^+ e^- \to \nu_e \bar \nu_e hh$ 
and $e^+ e^- \to hhhZ$ versus
the mass of light $CP$-even Higgs boson at $\sqrt{s}=$500 GeV
\vskip 5mm
Fig. 3 Total cross sections for the reactions $e^+ e^- \to hhZ$,
$e^+ e^- \to hhA$, $e^+ e^- \to \nu_e \bar \nu_e hh$ 
and $e^+ e^- \to hhhZ$ versus
the mass of heavy $CP$-even Higgs boson at $\sqrt{s}=$500 GeV
\vskip 5mm
Fig. 4 Two-body branching ratios of heavy $CP$-even Higgs boson
\vskip 5mm
Fig. 5 Two-body branching ratios of $CP$-odd Higgs boson

\newpage

\begin{figure}[t]
\begin{center}
{\def\chepscale{1.0} 
\unitlength=\chepscale pt
\SetWidth{0.7}      
\SetScale{\chepscale}
\scriptsize    
\begin{picture}(80,80)(0,0)
\Text(16.0,65.7)[r]{$e^-$}
\ArrowLine(16.5,65.7)(32.5,53.1) 
\Text(16.0,40.4)[r]{$e^+$}
\ArrowLine(32.5,53.1)(16.5,40.4) 
\Text(40.3,53.9)[b]{$Z$}
\DashLine(32.5,53.1)(48.5,53.1){3.0} 
\Text(65.0,65.7)[l]{$Z$}
\DashLine(48.5,53.1)(64.5,65.7){3.0} 
\Text(48.0,40.4)[r]{$h$}
\DashLine(48.5,53.1)(48.5,27.8){1.0}
\Text(65.0,40.4)[l]{$h$}
\DashLine(48.5,27.8)(64.5,40.4){1.0}
\Text(65.0,15.2)[l]{$h$}
\DashLine(48.5,27.8)(64.5,15.2){1.0}
\Text(40,0)[b] {diagr.1}
\end{picture} \ 
\begin{picture}(80,80)(0,0)
\Text(16.0,65.7)[r]{$e^-$}
\ArrowLine(16.5,65.7)(32.5,53.1) 
\Text(16.0,40.4)[r]{$e^+$}
\ArrowLine(32.5,53.1)(16.5,40.4) 
\Text(40.3,53.9)[b]{$Z$}
\DashLine(32.5,53.1)(48.5,53.1){3.0} 
\Text(65.0,65.7)[l]{$h$}
\DashLine(48.5,53.1)(64.5,65.7){1.0}
\Text(48.0,40.4)[r]{$Z$}
\DashLine(48.5,53.1)(48.5,27.8){3.0} 
\Text(65.0,40.4)[l]{$h$}
\DashLine(48.5,27.8)(64.5,40.4){1.0}
\Text(65.0,15.2)[l]{$Z$}
\DashLine(48.5,27.8)(64.5,15.2){3.0} 
\Text(40,0)[b] {diagr.2}
\end{picture} \ 
\begin{picture}(80,80)(0,0)
\Text(16.0,65.7)[r]{$e^-$}
\ArrowLine(16.5,65.7)(32.5,53.1) 
\Text(16.0,40.4)[r]{$e^+$}
\ArrowLine(32.5,53.1)(16.5,40.4) 
\Text(40.3,53.9)[b]{$Z$}
\DashLine(32.5,53.1)(48.5,53.1){3.0} 
\Text(65.0,65.7)[l]{$h$}
\DashLine(48.5,53.1)(64.5,65.7){1.0}
\Text(65.0,40.4)[l]{$h$}
\DashLine(48.5,53.1)(64.5,40.4){1.0}
\Text(65.0,15.2)[l]{$Z$}
\DashLine(48.5,53.1)(64.5,15.2){3.0} 
\Text(40,0)[b] {diagr.3}
\end{picture} \ 
\begin{picture}(80,80)(0,0)
\Text(16.0,65.7)[r]{$e^-$}
\ArrowLine(16.5,65.7)(32.5,53.1) 
\Text(16.0,40.4)[r]{$e^+$}
\ArrowLine(32.5,53.1)(16.5,40.4) 
\Text(40.3,53.9)[b]{$Z$}
\DashLine(32.5,53.1)(48.5,53.1){3.0} 
\Text(65.0,65.7)[l]{$Z$}
\DashLine(48.5,53.1)(64.5,65.7){3.0} 
\Text(48.0,40.4)[r]{$H$}
\DashLine(48.5,53.1)(48.5,27.8){1.0}
\Text(65.0,40.4)[l]{$h$}
\DashLine(48.5,27.8)(64.5,40.4){1.0}
\Text(65.0,15.2)[l]{$h$}
\DashLine(48.5,27.8)(64.5,15.2){1.0}
\Text(40,0)[b] {diagr.4}
\end{picture} \ 
\begin{picture}(80,80)(0,0)
\Text(16.0,65.7)[r]{$e^-$}
\ArrowLine(16.5,65.7)(32.5,53.1) 
\Text(16.0,40.4)[r]{$e^+$}
\ArrowLine(32.5,53.1)(16.5,40.4) 
\Text(40.3,53.9)[b]{$Z$}
\DashLine(32.5,53.1)(48.5,53.1){3.0} 
\Text(65.0,65.7)[l]{$h$}
\DashLine(48.5,53.1)(64.5,65.7){1.0}
\Text(48.0,40.4)[r]{$A^0$}
\DashLine(48.5,53.1)(48.5,27.8){1.0}
\Text(65.0,40.4)[l]{$h$}
\DashLine(48.5,27.8)(64.5,40.4){1.0}
\Text(65.0,15.2)[l]{$Z$}
\DashLine(48.5,27.8)(64.5,15.2){3.0} 
\Text(40,0)[b] {diagr.5}
\end{picture} \ 
}
\end{center}
\caption{  }
\end{figure}
.
\vskip 5cm

\newpage

\begin{figure}[t]
\begin{center}
\begin{picture}(8,8)
\put(-3,-1){\epsfxsize=13cm
         \epsfysize=10cm \leavevmode \epsfbox{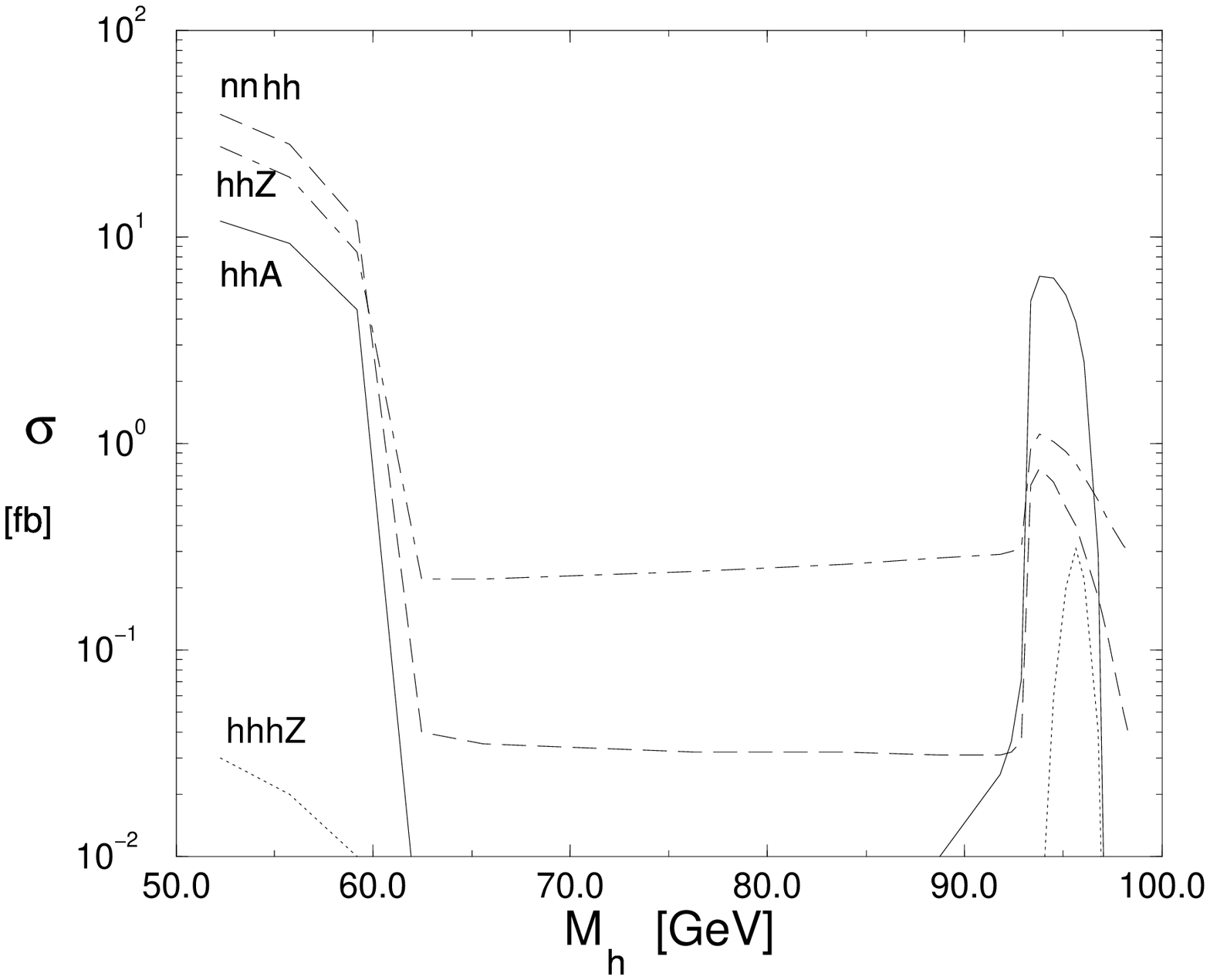}}
\end{picture}
\end{center}
\caption{  }
\end{figure}

\begin{figure}[b]
\begin{center}
\begin{picture}(8,8)
\put(-3,-1){\epsfxsize=13cm
         \epsfysize=10cm \leavevmode \epsfbox{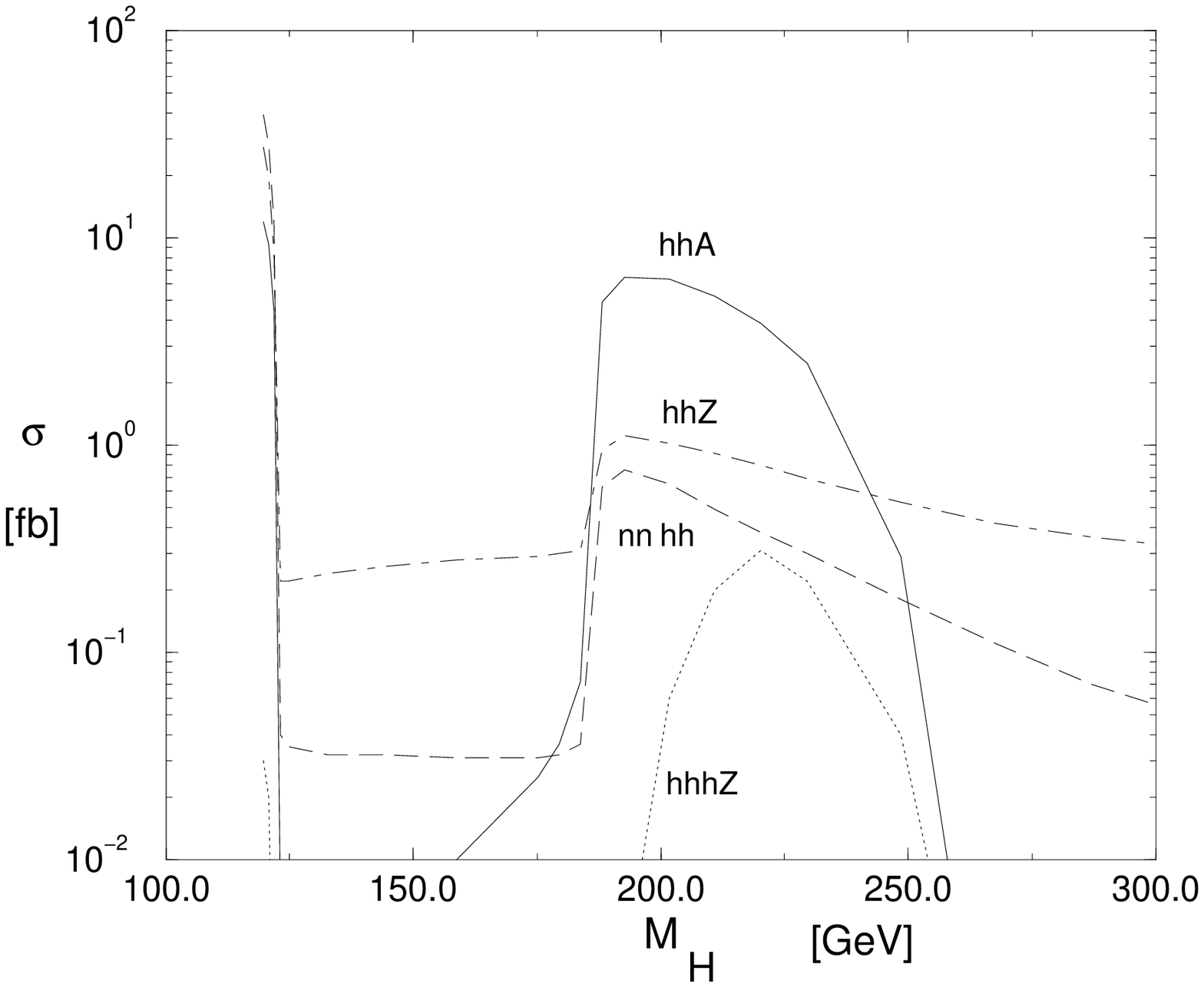}}
\end{picture}
\end{center}
\caption{  }
\end{figure}

\newpage

\begin{figure}[t]
\begin{center}
\begin{picture}(8,8)
\put(-3,-1){\epsfxsize=13cm
         \epsfysize=10cm \leavevmode \epsfbox{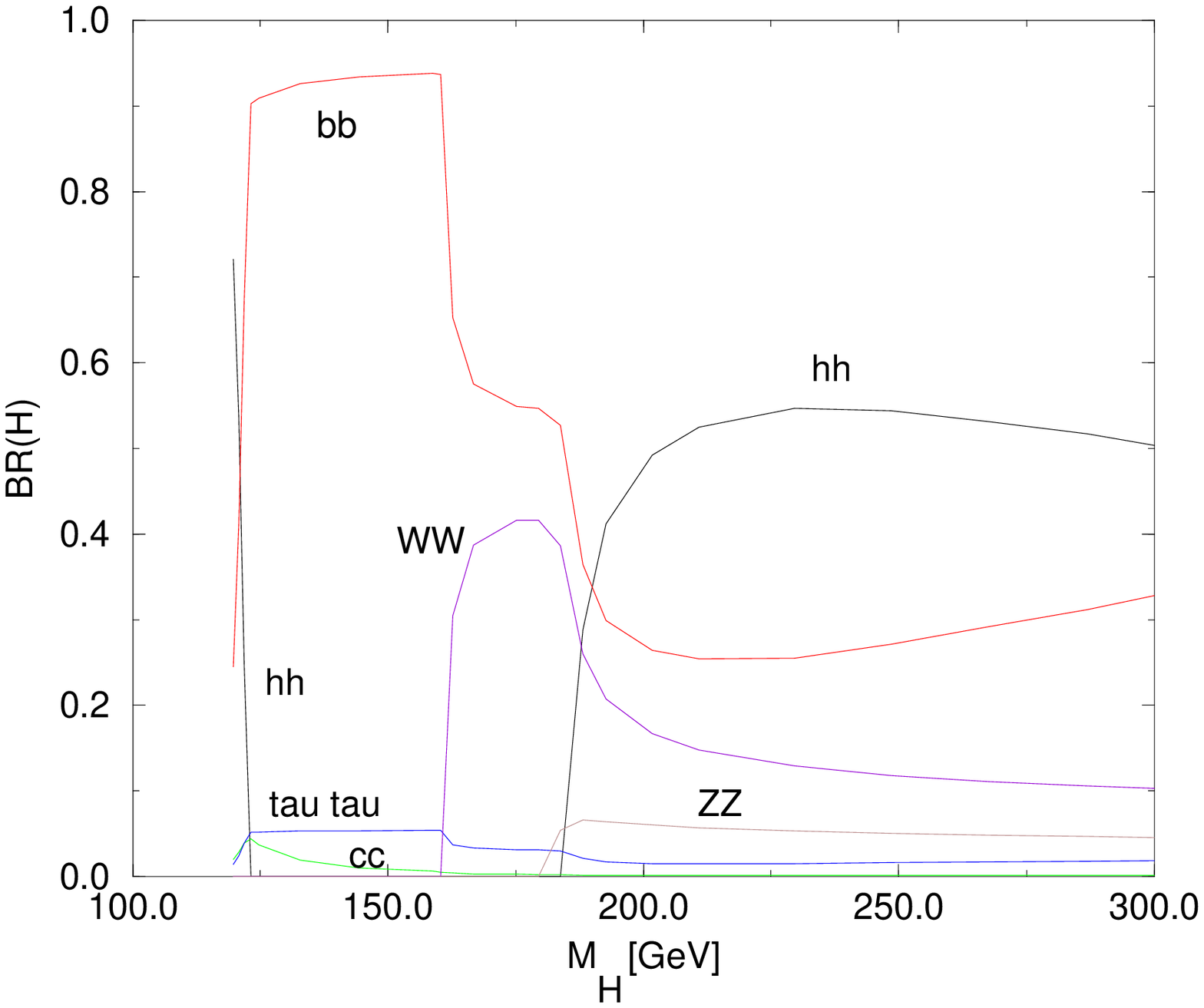}}
\end{picture}
\end{center}
\caption{  }
\end{figure}

\begin{figure}[b]
\begin{center}
\begin{picture}(8,8)
\put(-3,-1){\epsfxsize=13cm
         \epsfysize=10cm \leavevmode \epsfbox{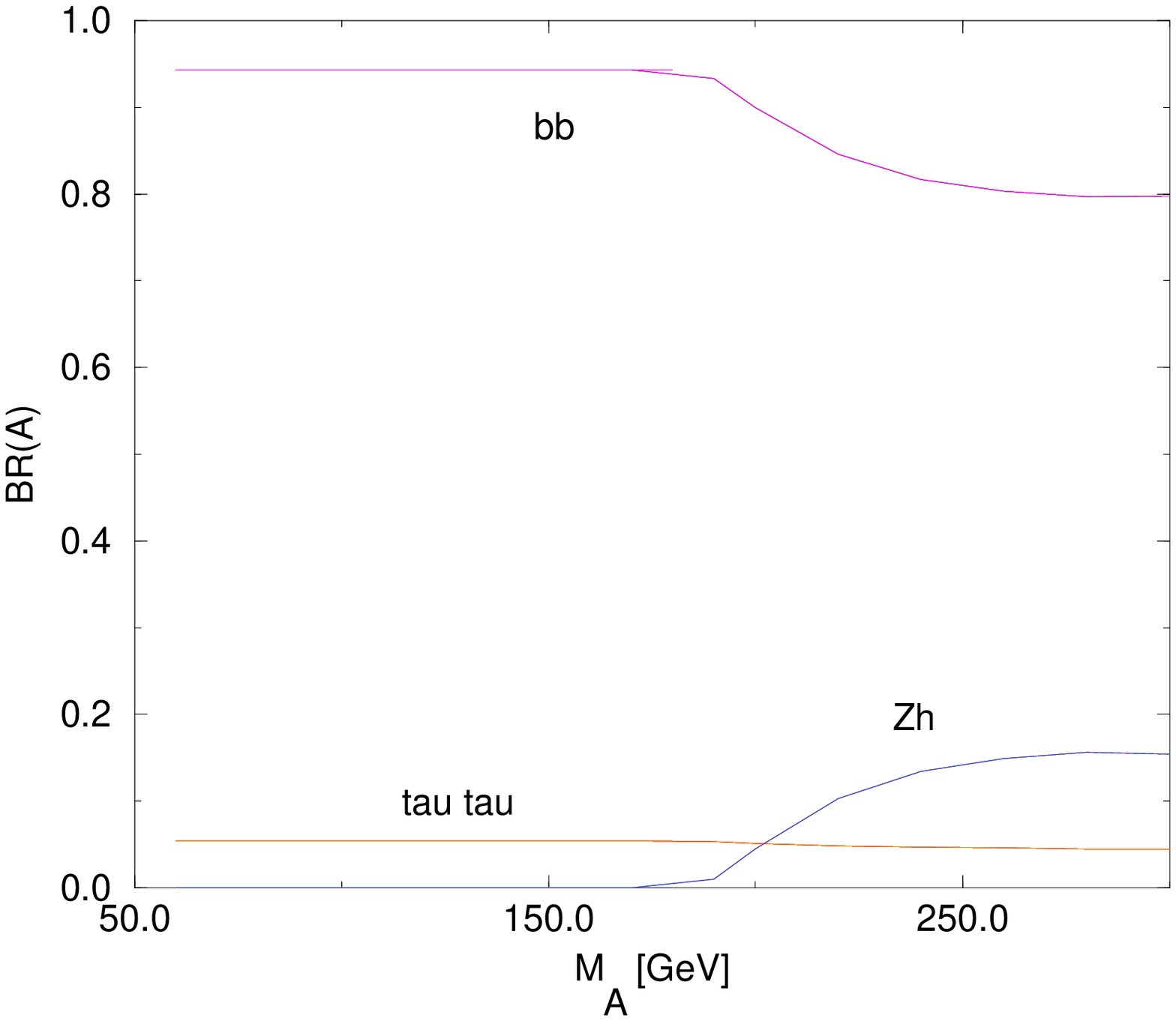}}
\end{picture}
\end{center}
\caption{  }
\end{figure}

\end{document}